\documentstyle[12pt,draft,epsf]{article}

\newcommand \be  {\begin{equation}}
\newcommand \ee  {\end{equation}}
\newcommand \bea {\begin{eqnarray} \nonumber }
\newcommand \eea {\end{eqnarray}}

\newcommand \gep {\epsilon}
\newcommand \get {\eta}

\newcommand \gs  {\sigma}

\newcommand \cN  {{\cal N}}

\def \t    {{\tau}}
\newcommand \ta   {{\tau'}}
\newcommand \tb   {{\tau''}}

\begin{document}

\title{Series Expansion of the Off-Equilibrium Mode Coupling
       Equations}
\author{ Silvio Franz$^{(a)}$, Enzo Marinari$^{(b)}$ and
         Giorgio Parisi$^{(c)}$\\[0.5em]
  {\small (a): Nordita and Connect}\\
  {\small \ \  Blegdamsvej 17, DK-2100 Copenhagen 0 (Denmark)}\\
  {\small (b): Dipartimento di Fisica and Infn, Universit\`a di Cagliari}\\
  {\small \ \  Via Ospedale 72, 09100 Cagliari (Italy)}\\
  {\small (c): Dipartimento di Fisica and Infn, Universit\`a di Roma
    {\em La Sapienza}}\\
  {\small \ \  P. A. Moro 2, 00185 Roma (Italy)}\\[0.3em]
   {\small \tt franz@nordita.dk}\\
   {\small \tt marinari@ca.infn.it}\\
   {\small \tt parisi@roma1.infn.it}\\[0.5em]}
\date{June 20, 1995}
\maketitle

\begin{abstract}
We show that computing the coefficients of the Taylor expansion of the
solution of the off-equilibrium dynamical equations characterizing
models with quenched disorder is a very effective way to understand
the long time asymptotic behavior. We study the $p=3$ spherical spin
glass model, and we compute the asymptotic energy (in the critical
region and down to $T=0$) and the coefficients of the time decay of
the energy.
\end{abstract}

\vfill

\begin{flushright}
  { \tt cond-mat/9506108 }
\end{flushright}
\newpage

It has been strongly stressed in the last period of time that the
non-equilibrium dynamics of glassy systems displays very interesting
phenomena. One of the most remarkable ones is {\em aging}
\cite{BOUCHA}. The dependence on time of measurable quantities does
not vanish when the time goes to $\infty$, and a true equilibrium is
never reached.

Let us give a simple example of an {\em aging} behavior.  In spin
systems the spin-spin correlation function can be defined as

\be
  C(t_w,t) \equiv N^{-1} \sum_i \gs_i(t_w) \gs_i(t_w+t) \ .
\ee
We can consider a system that has been kept at very high $T$ for times
smaller than $t=0$, and has been quenched to the measurement
temperature $T_0$ at time $t=0$. We wait a time $t_w$ after such a
quenching, and we measure spin-spin correlations starting at $t_w$.
The relevant point is that contrary to equilibrating systems, for
large $t_w$ and $t$ the dependence of $C$ on the waiting time $t_w$
does not disappear. In the simplest cases it is found that for
$\frac{t}{t_w} \ne 1$ $ C(t_w,t)$ can be written as

\be
C(t_w,t) = f({t \over t_w}),
\ee
where $f$ is not a constant function.
Even for asymptotically large times the time
translational invariance of the correlation functions is never
recovered.

Aging is quite a widespread phenomenon in short range models.  Aging
in disordered systems has the interesting peculiarity of being already
present in the mean field approximation. It is obvious that we can
have true, complete aging only in an infinite system. One should
investigate the dynamics of an infinite system. In this case one can
write a closed set of dynamical equations(\cite{CRHOSO,CUKU}) which
are the appropriate generalization of the famous mode-mode coupling
equation for glasses.  Unfortunately these equations are coupled
integral equations in two variables and the analytic solutions found
so far \cite{CUKU,FRAMEZ,CUKUSK,FRAPAR}, although they give a coherent
picture of aging, do not allow to study the time dependence of the
different quantities. It is then highly desirable to have efficient
methods to integrate the equations numerically.

An example of these equations for the spherical spin model with a
$p$-spin interaction in the case $p=3$ is given by a set of non-linear
integro-differential equations (\cite{CRHOSO,CUKU}), which are
obtained by transforming eqs. (\ref{E_CEQ},\ref{E_REQ}) back to their
constrained form (see the discussion in the following).  The same
equations also appear in the off-equilibrium dynamics of the
Amit-Roginsky model \cite{AR}, where although quenched disorder is
absent \cite{FH} the Mode Coupling Approximation gives exact results.

It is possible to study such equations numerically \cite{FRAMEZ}, but
it is rather difficult to study the solution for large values of the
time. The main problem is of the amount of information which needs to
be stored, since if the discretization step in time is $\gep$ the
number of numbers needed to code the correlation function is $L^2$
where $L= t/\gep$.

The dynamical equations can be derived starting from the usual
Langevin equation for the $p$-spin model

\be
  \dot{\gs}_i(t) = - \mu(t)\gs_i(t) + \sum_{i_2<...<i_p}^{1,N}
  J_{i,i_2,...,i_p}\gs_{i_2}(t) ... \gs_{i_p}(t) + \get_i(t)\ ,
\ee
where the $J$ couplings are quenched Gaussian random variables, and
$\eta$ is a white noise with covariance $2T$.  For all times $t$ the
value of $\mu(t)$ is chosen such to implement the spherical constraint

\be
 \sum_{i=1}^N \gs_i(t)^2=N\ .
\ee
For mainly technical purposes, and some extetical considerations, we
find more suitable to switch to a formulation where the spins $s_i(t)$
are unconstrained. All what we have have done here can be repeated,
with essentially the same degree of complexity, in the formulation
where the spins are the original constrained variables.

If we consider the transformation

\be
  s_i(t) = \cN(t) \gs_i(t)\ ,
\ee
with

\be
  \cN(t) = e^{ \int_0^t dt' \mu(t')}\ ,
\ee
the dynamical equations become

\be
  \cN(t)^{p-2} \dot{s}_i(t) =   \sum_{i_2<...<i_p}^{1,N}
  J_{i,i_2,...,i_p}s_{i_2}(t) ... s_{i_p}(t) +
  \cN(t)^{p-1} \get_i(t)\ .
\ee
Defining an effective time $\tau$ from the relation

\be
  dt = \cN^{p-2} d\tau\ ,
\ee
we get that

\be
  \protect\label{E_DIN}
  \frac{\partial s_i(\tau)}{\partial\tau} =
  \sum_{i_2<...<i_p}^{1,N}
  J_{i,i_2,...,i_p}s_{i_2}(\tau) ... s_{i_p}(\tau) +
  \xi_i(\tau)\ ,
\ee
where the $\xi_i$ are a gaussian noise with zero mean and variance

\be
  \protect\label{E_CSICSI}
  \overline{\xi_i(\tau)\xi_j(\tau')} = 2\  T\  \cN^p(\tau)
  \delta(\tau-\tau') \delta_{i,j}\ .
\ee
The unconstrained system of the $s$ variables is kept at a temperature

\be
  T_{eff}(\tau) \equiv T \cN^p(\tau)\ ,
\ee
in order to keep the original, constrained variables, at a fixed,
time independent temperature $T$.

 From the previous equations (\ref{E_DIN},\ref{E_CSICSI})
after standard manipulations we obtain a set of
closed dynamical equations for the correlation function

\be
  C(\tau,\tau') \equiv \langle s_i(\tau) s_i(\tau') \rangle\ ,
\ee
and for the response function

\be
  R(\tau,\tau') \equiv \frac{\delta \langle s_i(\tau)\rangle}
                       {\delta \xi_i(\tau')}\ .
\ee
For $\tau > \tau'$ they are respectively

\bea
  \nonumber
  \frac{\partial}{\partial\t} C(\tau,\ta) &=&
  \frac12 p(p-1) \int_0^\tau  d\tau'' C^{p-2}(\tau,\tb)
  R(\t,\tb) C(\tb,\ta)  \\
  &+& \frac12 p \int_0^\ta d\tb C^{p-1}(\t,\tb) R(\ta,\tb)\ ,
  \protect\label{E_CEQ}
\eea
and

\be
  \frac{\partial}{\partial\t} R(\tau,\ta) =
  \frac12 p(p-1) \int_\ta^\t  d\tau'' C^{p-2}(\tau,\tb)
  R(\t,\tb) R(\tb,\ta)\ .
  \protect\label{E_REQ}
\ee
For equal times $R=1$, and $C$ verifies

\bea
  \nonumber
  \frac12 \frac{\partial}{\partial\t} C(\t,\t) &=&
  \frac12 p^2 \int_0^\tau  d\tb C^{p-1}(\t,\tb)
  R(\t,\tb) \\
  &+& T C(\t,\t)^{\frac{p}{2}}\ ,
\eea
where we used the fact that

\be
  \cN(\tau) = \sqrt{C(\t,\t)}\ .
\ee
At the end of the day we have rewritten our constrained system as an
unconstrained system, kept at a time-dependent temperature
$T_{eff}(\tau)$. We stress that this transcription
has been allowed by the fact that our
Hamiltonian is homogeneous. We will
discuss the system in such an unconstrained formulation. We remind
again the reader that all our results can be easily reobtained in the
constrained formalism, where one can also deal with non-homogeneous
Hamiltonians as the ones in \cite{FRAMEZ}.

The method we suggest and employ here to study the asymptotic behavior of the
dynamic equations (\ref{E_CEQ}) and (\ref{E_REQ}) is based on computing the
coefficients of the Taylor expansion of $C$ and $R$, defined by

\bea \nonumber
 C(\tau_1,\tau_2) &=& \sum_{k,j} c_{k,j} \tau_1^k \tau_2^j\ ,\\
 R(\tau_1,\tau_2) &=& \sum_{k,j} r_{k,j} \tau_1^k \tau_2^j\ .
\eea
We are always assuming $\tau_1>\tau_2$.  Equations (\ref{E_CEQ}) and
(\ref{E_REQ}) translate in two coupled iterative relations for the
Taylor coefficients $c$ and $r$. Knowing the lower order coefficients
we can determine the higher orders one. The initial conditions are
$c_{0,0}=1$ and $r_{0,0}=1$.  We determine in turn the coefficients
$c_{k,j}$ and $r_{k,j}$ with $i+j=\omega$ for $\omega=2$, $3$,
$4...$. It takes a few hours of a RISC workstation to go
up to $\omega=100$ (even if we analyze eventually only
coefficients going up to the order $48$, see later). We repeat the
procedure for different values of the temperature $T$. The complexity
of the computation increases as $\omega^3$.

In this note, as an example, we give the results of the computation of
the energy, which happen to be in agreement with the theory
\cite{CRHOSO,CUKU}.  In a forthcoming publication we will address the
problem of verifying more striking issues of the theory as the
asymptotic violation of time translation invariance and of the
fluctuation dissipation theorem.

We start the analysis from the Taylor expansion for $C(\tau,\tau)$. If we
define

\be
  c_k^{(d)} \equiv \sum_{j=0}^{k} c_{k-j,k}\ ,
\ee
we have that

\be
  C(\tau,\tau) = \sum_{l=0}^\infty c_l^{(d)}\tau^l\ .
\ee
In the self-explaining formalism we will use in the following we will be
working on the Taylor coefficients

\be
  \{ C^{(d)}(\tau) \}_k\ .
\ee
We have computed the coefficients of the series expansions which follow by
using a simple program written in $C$-language. We have at first computed the
coefficients

\be
  \{\cN(\tau)\}_k = \{ C^{(d)}(\tau)^{\frac12} \}_k =
  \{ e^{ \int_0^{t(\tau)} dt'\mu(t')}  \}_k\ .
\ee
The coefficients of $\mu$ as a function of the {\em unconstrained time} $\tau$
are easily computed as

\be
  \{ \mu(\tau)  \}_k =
  \{ \frac{1}{\cN}\frac{d\cN}{dt}  \}_k =
  \{ \frac{1}{\cN}\frac{d\cN}{d\tau} \frac{d\tau}{dt} \}_k =
  \{ - \frac{d\cN^{-1}}{d\tau} \}_k \ ,
\ee
and for the {\em constrained time} $t$ as a function of the
{\em unconstrained time}  $\tau$

\be
  \{  t(\tau )\}_k = \{ \int_0^td\tau \cN(\tau)\}_k = \frac{\cN_{k-1}}{k}\ .
\ee
By inverting and composing these series expansions we eventually obtain

\be
  \{ \mu(t) \}_k\ ,
\ee
basically the energy as a function of the {\em constrained time} $t$. It is
useful to define at last the Taylor expansion of the function

\be
  \{ \beta(t)  \}_k \equiv \{ t \frac{\mu''(t)}{\mu'(t)}  \}_k\ .
\ee
Under the assumption that, for $t\to\infty$,

\be
  \mu(t) \to \mu_\infty - A t^{-\alpha}\ ,
\ee
one has that

\be
  \beta(t) \to - ( \alpha + 1 )   \ .
\ee
Different techniques may be used to extrapolate the function
$\beta(t)$ to $t=\infty$. After a series of tests we have found
convenient to use Pad\`e approximants.
We have used the diagonal Pad\`e approximants of order $48$ to check
this behavior\footnote{Lower order Pad\`e approximants have been used
to check stability.}. We compute $\alpha$ from $\beta(\tilde{t})$,
with an high enough value of $\tilde{t}$. having our best estimate for
$\alpha$ we compute $\mu(\tilde{t})$ and the extrapolated value
$\mu(\tilde{t}=\infty)$.
Once $\mu(t)$ is known, the energy $E(t)$ is obtained by the simple
relation $\mu(t)=T-3 E(t)$ \cite{CRHOSO}.

Let us start with a brief summary of our results.  Our method confirms
that the asymptotic dynamic energy, $E(T)$ is correctely predicted by the
theory.  We find in addition the remarkable result that the value of
the asymptotic dynamic energy of the dynamics at zero temperature is
the smooth limit of the values for $T\ne 0$.  This does not seem to be
true in other spin-glass systems as for example the SK model
\cite{opper}.  We also get a very good estimate for the exponent
$\alpha$ for $T<T_D=0.5$. These results have been obtained with very
small computational effort.

\begin{figure}
\epsfxsize=400pt\epsffile{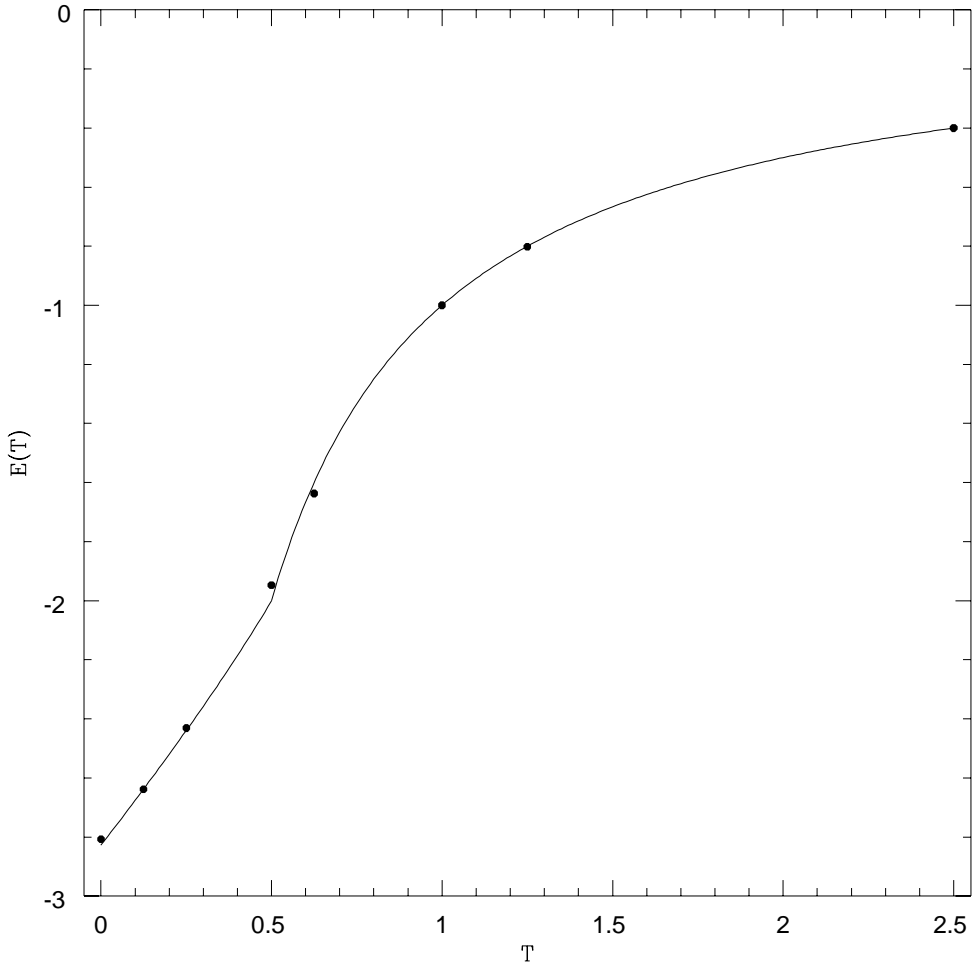}
  \caption[0]{\protect\label{F_E1}
     Theoretical result for the asymptotic dynamic energy $E(T)$ (solid
curve) and our values (filled dots). The dynamic critical temperature
is $T_D=\frac12$. In the high $T$ region $E(T)=\frac{1}{T}$.}
\end{figure}

In figure (\ref{F_E1}) we compare the theoretical result for the
asymptotic value of the dynamic energy to our findings. Here the
critical temperature is obtained by solving the equation

\be
 \frac{2}{T_D^2} q_D (1-q_D)^2 = 1,\ q_D \equiv q(T_D) = \frac12
\ee
that gives
$T_D=\frac12$, and the dynamic energy is defined as

\be
  -\frac{1}{T} (1+q^2(1-2q))\ .
\ee
The agreement is very good at low $T$. The convergence (in time)
becomes slower when approaching the critical point, and exactly at
$T_D$ we have the higher discrepancy from the exact value (of more or
less $2\%$). In the high $T$ phase the convergence to the asymptotic
result becomes very fast again.

\begin{figure}
\epsfxsize=400pt\epsffile{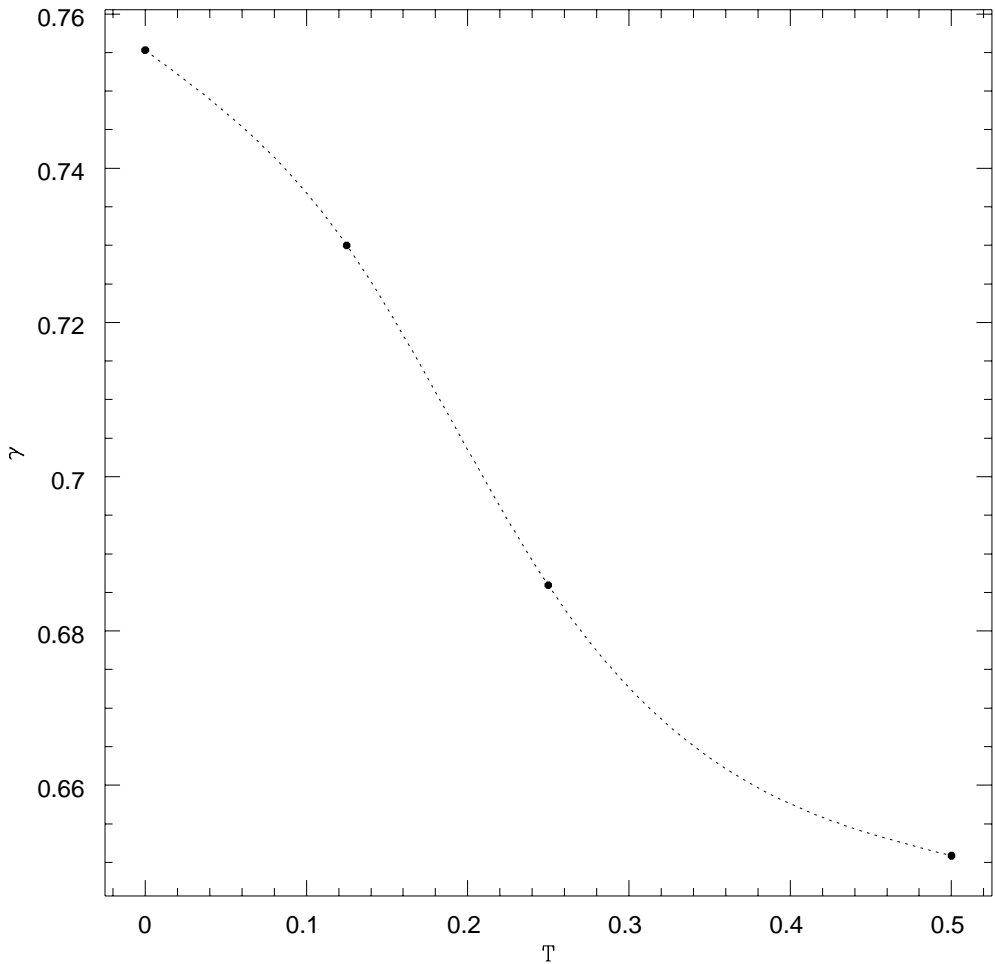}
  \caption[0]{\protect\label{F_GAMMA}
  $\alpha$ as a function of $T$. In these units the dynamic transition
temperature $T_D=\frac12$. The smooth curve is just used for
presenting the data, and does not have any physical meaning of a best
fit.}
\end{figure}

In figure (\ref{F_GAMMA}) we plot the exponent $\alpha$ we have
estimated from Pad\`e approximants as a function of $T$. We do not
have a precise estimate of the errors (this is a known drawback of the
Pad\`e approach), but judging from the dispersion of different
diagonal approximants they appear to be of the order of magnitude of a
few percent. If we try to estimate $\alpha$ for
$T>T_D$, in the high temperature phase, where we expect the
correlations to decay exponentially, we find a value that starts to
increase with $T$ and dramatically explodes for high $T$. The
relatively large discrepancy of the extrapolated energy value close to
$T_D$ makes us suspicious about the possible presence of tricky
confluent singularities. Obviously such a problem would reflect itself
in the value we are estimating for $\alpha$.

\begin{figure}
\epsfxsize=400pt\epsffile{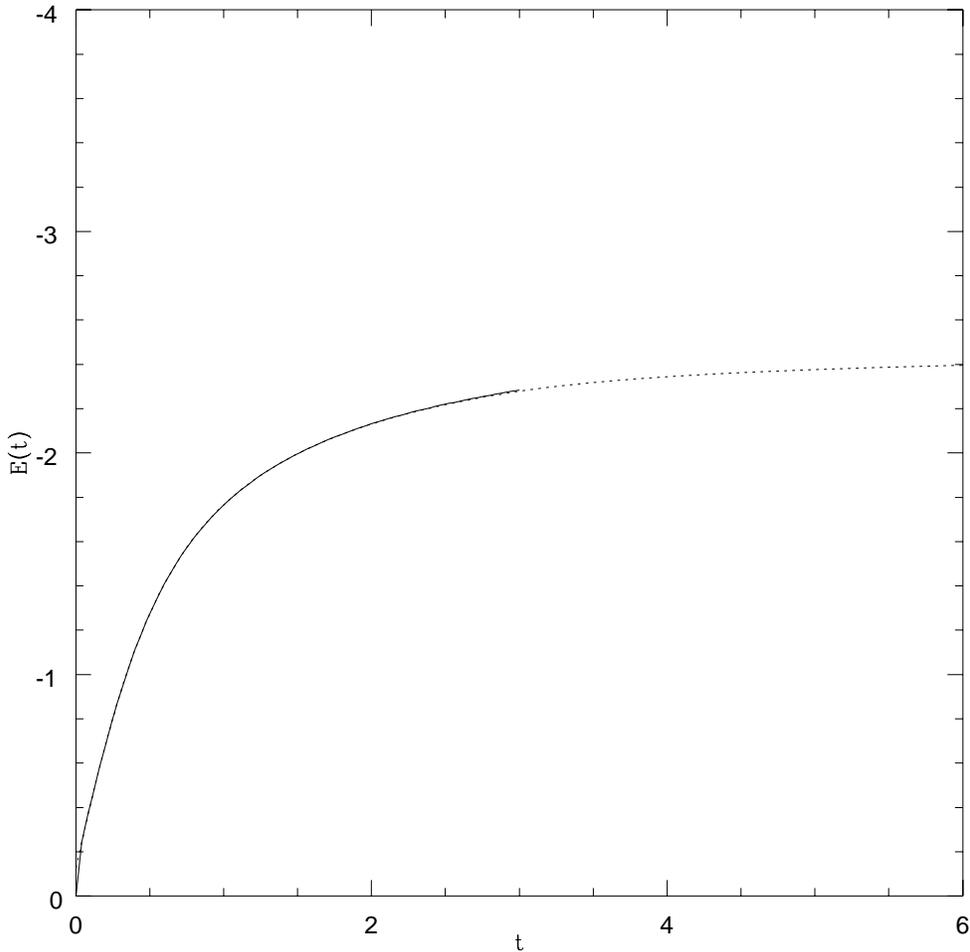}
  \caption[0]{\protect\label{F_E2}
     Result for the asymptotic dynamic energy $E(t)$ from the
direct
numerical integration of the dynamical equations (solid
curve) and our result as a function of time (dotted line).
here $T=\frac{T_D}{4}$, i.e. $T=0.125$ with our normalization.}
\end{figure}

The fact that our method works very well is also clear from figure
(\ref{F_E2}), where we compare the results of our expansion to the
result from the direct
numerical integration of the dynamical equations. We
have a perfect matching, but, as we discussed,
 the  numerical integration is
limited to small times, and it is difficult to extract from there
reliable asymptotic values for the coefficients.

Summing up, we have seen that the numerical integration of
the off-equilibrium mode coupling equations can effectively
be performed computing the coefficients of the Taylor expantion
of the correlation and the response functions. As an example of
application we have computed the energy as a function of time
and temperature. In the low temperature phase (and
down to zero temperature) we find power
law decay to the value predicted by the mean field theory.

\end{document}